# Large anomalous Hall effect in a hexagonal ferromagnetic Fe$_5$Sn$_3$ single crystal

Hang Li[1,4], Bingwen Zhang[2], Jinghua Liang[3], Bei Ding[1,4], Jie Chen[1,4], Jianlei Shen[1,4], Zefang Li[1,4], Enke Liu[1,5], Xuekui Xi[1], Guangheng Wu[1], Yuan Yao[1], Hongxin Yang[3,4] and Wenhong Wang[1,5*]

[1]Beijing National Laboratory for Condensed Matter Physics, Institute of Physics, Chinese Academy of Sciences, Beijing 100190, China
[2]Fujian Provincial Key Laboratory of Functional Marine Sensing Materials, Center for Advanced Marine Materials and Smart Sensors, Minjiang University, Fuzhou 350108, China
[3]Ningbo Institute of Materials Technology and Engineering, Chinese Academy of Sciences, Ningbo 315201, China
[4]University of Chinese Academy of Sciences, Beijing 100049, China
[5] Songshan Lake Materials Laboratory, Dongguan, Guangdong 523808, China

## Abstract

In this paper, we report an experimental observation of the large anomalous Hall effect (AHE) in a hexagonal ferromagnetic Fe$_5$Sn$_3$ single crystal with current along the b axis and a magnetic field normal to the bc plane. The intrinsic contribution of the anomalous Hall conductance $\sigma_{AH}^{int}$ was approximately 613 $\Omega^{-1}$ cm$^{-1}$, which was more than 3 times the maximum value in the frustrated kagome magnet Fe$_3$Sn$_2$ and nearly independent of the temperature over a wide range between 5 and 350 K. The analysis results revealed that the large AHE was dominated by a common, intrinsic term, while the extrinsic contribution, i.e., the skew scattering and side jump, turned out to be small. In addition to the large AHE, it was found the types of majority carriers changed at approximately 275 and 30 K, consistent with the critical temperatures of the spin reorientation. These findings suggest that the hexagonal ferromagnetic Fe$_5$Sn$_3$ single crystal is an excellent candidate to use for the study of the topological features in ferromagnets.

---

*Authors to whom correspondence should be addressed. Email: wenhong.wang@iphy.ac.cn

# I. INTRODUCTION

The anomalous Hall effect (AHE), discovered by E. H. Hall a century ago [1], has attracted generous interest from both the physics and materials communities because of its controversial physical origin [2-5] and wide practical application [6]. In contrast to the electromagnetic origin of the ordinary Hall effect, $\rho_{yx}^{OH} = R_0\mu_0 H$, the AHE contribution to the total Hall resistivity, $\rho_{yx} = \rho_{yx}^{OH} + \rho_{yx}^{AH}$, is determined solely by the spontaneous magnetization, $\rho_{yx}^{AH} = R_s M$ [7-9]. In the formula, $R_0$ and $R_s$ are the ordinary and anomalous Hall coefficients, respectively. The magnetization M is a function of the external field H. On the theoretical side, Karplus and Luttinger (K-L) proposed the intrinsic mechanism that considered the $\rho_{yx}^{AH}$ was quadratic in the longitudinal zero-field resistivity $\rho_{xx}$ ($\rho_{yx}^{AH} \propto \rho_{xx}^2$) [2]. The K-L mechanism is understood as the effect of the spin-orbit interaction of polarized conduction electrons at an external electrical field. With this in mind, Smit [3,4] and Berger [5] identified two basic extrinsic mechanisms, skew-scattering and side-jump, which were linear ($\rho_{yx}^{AH} \propto \rho_{xx}$) and quadratic ($\rho_{yx}^{AH} \propto \rho_{xx}^2$) for $\rho_{xx}$, respectively. These mechanisms are commonly understood to be a result of asymmetric scattering from the spin-orbit interaction acting on conducting electrons or impurities. Recently, the intrinsic mechanism (K-L) has been considered to be linked to the Berry curvature of electronic band structure in momentum space, which induces an anomalous velocity perpendicular to the electric field [10,11].

On the experimental side, a few materials with complex magnetic structures have been found to have non-trivial band structures, such as the non-collinear antiferromagnet $Mn_3Sn$ [12,13], frustrated kagome ferromagnet $Fe_3Sn_2$ [14,15], ferromagnetic Weyl semimetal $Co_3Sn_2S_2$ [16-18], and two-dimensional ferromagnet $Fe_3GeTe_2$ [19]. The non-trivial band structure always couples with a large Berry curvature [20] that acts like a fictitious magnetic field and generates a large intrinsic AHE [10,11]. Therefore, to some extent, a large AHE has become a symbol of searching for topological materials. The intrinsic magnetic topological materials are the ideal candidates with which to explore the relationship between magnetism and topological band structures, which has propelled researchers to find more intrinsic magnetic topological materials.

In this paper, a large AHE in a hexagonal ferromagnetic $Fe_5Sn_3$ single crystal, which was mostly induced by the intrinsic mechanism, is reported. As shown in Figs. 1(a) and 1(b), $Fe_5Sn_3$ single crystals crystallized in a hexagonal crystal structure with the space group $P6_3/mmc$ (a=b=4.224 Å, c=5.222 Å). The Fe and Fe-Sn layers were arranged alternately along the c axis. Two different sites for the Fe atoms were located at Fe-I (0, 0, 0) and Fe-II (1/3, 2/3, 1/4) and the occupancies of the Fe-I and Fe-II are 1 and 0.67, respectively. Furthermore, the Fe-II sites always formed vacancies or disorder that were related to the synthesis temperatures and they further affected the Curie temperature $T_C$ and the saturation magnetization $M_s$ of the $Fe_5Sn_3$ [21,22]. The easy magnetization axis of the $Fe_5Sn_3$ tended to be the b axis [23]. However, the detailed magnetic structure of $Fe_5Sn_3$ awaits further study.

## II. EXPERIMENTAL DETAILS

Single crystals of $Fe_5Sn_3$ were synthesized by the Sn-flux method with a molar ratio of Fe:Sn =1:17. Fe (purity 99.95%) and Sn (purity 99.99%) grains were mixed and placed in an alumina crucible. The crucible was then sealed in a tantalum tube under a partial Argon atmosphere. Finally, the tantalum tube was sealed in a quartz tube to avoid oxidation. The quartz tube was placed in a furnace and kept at 1150 °C for two days to obtain a homogeneous metallic solution, and then the quartz tube was cooled from 910 °C to 800 °C at a rate of 1.5 °C/h and kept at 800 °C for two days to reduce the defects in the single crystals. To obtain isolated single crystals, the quartz tube was moved quickly into the centrifuge to separate the excess Sn flux at 800 °C. As shown in the inset of Fig. 1(c), the free-growing single crystal of $Fe_5Sn_3$ appeared as a bar (approximately 2–5 mm long) with a hexagonal section. X-ray diffraction (Fig. 1(c)) and high-resolution morphology (Fig. 1(d)) examinations were conducted at room temperature with a Bruker D2 X-ray machine with Cu $K_\alpha$ radiation (λ=1.5418 Å) and a scanning transmission electron microscope (JEOL ARM200F), respectively. To confirm the orientations of the single crystal planes, the single crystal X-ray diffraction (see Fig. S1 of the Supplementary Material [24]) was also measured. These results strongly suggested that the rectangular plane was the bc plane and the long side was the c axis. Energy dispersive X-ray spectroscopy (EDS) measurements showed that the average ratio of iron in the samples was approximately 61.67% (see Table S1 and Fig. S2 [24]), which was less than the standard stoichiometric ratio. The magnetization and electrical transport were measured in a Quantum Design MPMS-XL and PPMS-9, respectively. To measure the electrical transport, the single crystals were cut into a rectangle with dimensions of 1.5×0.5×0.1 mm and a standard four-probe method was applied. To eliminate the influence of the voltage probe misalignment, the longitudinal ($\rho_{xx}$) and transverse ($\rho_{yx}$) resistivities were measured for both positive and negative fields ($\rho_{xx}(\mu_0 H) = (\rho_{xx}(+\mu_0 H) + \rho_{xx}(-\mu_0 H))/2$ and $\rho_{yx}(\mu_0 H) = (\rho_{yx}(+\mu_0 H) - \rho_{yx}(-\mu_0 H))/2$).

## III. RESULTS AND DISCUSSION

Figure 2(a) shows the zero field cooling curves (ZFC) at 0.5 T with the magnetic fields H parallel to the b and c axis. It is clearly shown in the figure that the easy magnetization axis of the single crystal tended to be the b axis and the Curie temperature $T_C$ was approximately 600 K, consistent with early reports [22,23]. The inset of Fig. 2(a) shows the low temperature details of the magnetization with H along the c axis at 0.5 T. Two obvious kink points appeared at approximately 50 and 125 K (the more detailed M-T curves with H along the c axis are shown in Fig. S3 [24]). The uniaxial magnetic anisotropy coefficient $K_u = \mu_0 M_s H_k/2$ decreased with increasing temperature, shown in Fig. 2(b), suggesting that the easy axis gradually rotated from the ab plane to the c axis with increasing temperature. The $H_k$ term in the formula is the anisotropy effective field that is defined as the critical field above the difference in magnetization between the two magnetic field directions (H//b axis and H//c axis) becoming smaller than 2%. Remarkably, there were also two clear kink points at 250 and 125 K. However, the curves of the heat capacity $C_p$ (10–100 K) and the differential scanning calorimetry (DSC, 100–400 K), shown in Fig. S5 [24], did not exhibit any obvious peaks for the heating and cooling processes. Therefore, these kink points of the M-T and $K_u$-T curves near 50, 125, and 250 K could all be induced by the continuous rotation of the easy axis from the ab plane to the c axis with increasing temperature rather than the crystallographic phase transition. However, a full accounting of the magnetic structure awaits further analysis. The inset of Fig. 2(b) shows the magnetization curves with H along the b axis and c axis at 300 K (more detailed M-H curves are shown in Fig. S4 [24]). It is clearly shown in the figure that the saturation fields along the two axes were approximately 0.2 T (b axis) and 1.5 T (c axis).

The temperature dependence of the longitudinal resistivity $\rho_{xx}$ along the b axis for zero field and 5 T are shown in Fig. 2(c). The curves of $\rho_{xx}$ exhibited typical metallic behavior at the zero field and at 5 T. There was a clear slope change of the zero field $\rho_{xx}$ curve near 50 K and it was conspicuously suppressed at 5 T, which also implied that the kink points of M-T and $K_u$-T curves near 50 K were induced by the magnetic structure transition. The inset shows the curves of $d\rho_{xx}/dT$ at zero field. The slopes of $\rho_{xx}$ appeared to have an obvious change below 125 K and the slopes reached a maximum at approximately 30 K, which was consistent with the changes of magnetization near these temperatures. The magnetoresistance (MR = $(\rho_{xx}(\mu_0 H) - \rho_{xx}(0)) \times 100\%/\rho_{xx}(0)$) is shown in Fig. 2(d). It is clearly shown in the figure that the sign of the MR was opposite at low and high temperatures and the turning point was approximately 50 K (more detailed MR curves are shown in Fig. S6(c) [24]). The negative MR was obviously induced by spin-dependent scattering in the magnetic systems [25] and the positive MR may relate to the complex magnetic structure [26] at the low-temperature zone in $Fe_5Sn_3$. The sign change was obviously associated with the transformation of the magnetic structure with temperature, which was similar to the situation in $Fe_3Sn_2$ [27].

Figure 3(a) shows the magnetization curves M-H with H normal to the bc plane at various temperatures. The M-H curves displayed typical soft ferromagnet behavior with a coercive field that was almost zero and a saturation field of approximately 1 T, which was between the values at H along the b axis and c axis. The saturation magnetization was approximately 2.29$\mu_B$/Fe at 5 K, which was slightly larger than the saturation magnetization in early reports [21,23,28]. The discrepancy arose from the different vacancies or the disorder of Fe-II atoms in the samples that were synthesized at different temperatures [22]. The data for the Hall resistivity $\rho_{yx}$ with the current along the b axis and H normal to the bc plane are shown in Fig. 3(b). The magnetic field dependence curves of $\rho_{yx}$ appeared to have a typical AHE for the Fe$_5$Sn$_3$, and more than one sample was measured. The more detailed magnetization and electrical transport data are shown in Figs. S4, S6, and S7 [24].

To further study the AHE of Fe$_5$Sn$_3$, a detailed analysis was made, the results of which is shown in Figs. 4(a)–4(d). Above the saturation field (approximately 1 T), the M was replaced by the saturation magnetization M$_s$ and the anomalous hall resistivity $\rho_{yx}^{AH}$ became a constant. With a linear fit of the $\rho_{yx}$ at the high-field region (above 2 T), which is shown by the red dashed line in Fig. 3(b), the slopes and intercepts corresponded to R$_0$ and the anomalous Hall resistivity $\rho_{yx}^{AH} = R_s M_s$. The values of M$_s$ were taken from the magnetization curves at 5 T. The curve of R$_0$-T is shown in Fig. 4(a). The signs of R$_0$ represent the types of majority carriers. An interesting phenomenon was that the signs of R$_0$ exhibited obvious changes at approximately 275 and 30 K in Fe$_5$Sn$_3$. The signs of R$_0$ were positive when the temperature was above 275 K or below 30 K, which meant that the majority carriers were electrons. The majority carriers changed into holes between 30 and 275 K. The inset of Fig. 4(a) shows the carrier concentrations (n=1/|e|R$_0$), which also appeared to have two local maximum points at 275 and 30 K. R$_0$ did not change its sign near 125 K, but the R$_0$ values appeared to have a local extreme value. The transformation temperatures showed a small discrepancy for different samples (see Fig. S7(c) [24]), which was because of the slight difference of the vacancies or the disorder in the site of Fe-II [22]. The facts described above are reminders of the continuous rotation of the easy axis, as shown in Figs. 2(a) and 2(b), which was also similar to the situations of the 2H metallic transition-metal dichalcogenides [29-31] and the film of the metallic oxides CaRuO$_3$ and SrRuO$_3$ [32]. The unconventional relationship between R$_0$ and T may have been caused by a tight correlation between the magnetic structure and the Fermi surface in Fe$_5$Sn$_3$. Figure 4(b) shows the temperature dependence of R$_s$, which increases monotonically with temperature. The inset shows the scaling coefficient $S_H = R_s/\rho_{xx}^2$, which was a material-specific scale factor [33,34].

To help understand the origin of the AHE in the Fe$_5$Sn$_3$, the function dependencies of $\log\rho_{yx}^{AH}$ and $\log\rho_{xx}$ are shown in Fig. 4(c). The slope of the linear fitting α was equal to 2.06, which meant that the AHE of Fe$_5$Sn$_3$ could be from the K-L mechanism [2] or the side jump mechanism [5], because both showed the same quadratic relationship between $\rho_{yx}^{AH}$ and $\rho_{xx}$ [35]. To further ascertain the mechanism of the AHE, a recently discovered method [36-38] was used to separate the intrinsic and extrinsic contributions, as shown in Fig. 4(d). The inset of Fig. 4(d) shows the scaling behavior of $\rho_{yx}^{AH}$ versus $\rho_{xx}(0\ T)$. The red curve in the figure represents the fitting equation $\rho_{yx}^{AH} = \alpha\rho_{xx0} + \beta\rho_{xx0}^2 + \gamma\rho_{xx}^2$, where $\rho_{xx0}$ is the residual resistivity (the value of $\rho_{xx}$ at 2 K was used), and α, β, and γ are the coefficients of the skew scattering, side-jump, and intrinsic terms, respectively. As shown in Fig. 4(d), a large intrinsic value ($\sigma_{AH}^{int}$~613 Ω$^{-1}$ cm$^{-1}$) was obtained for the anomalous Hall conductance (AHC) in Fe$_5$Sn$_3$, which was more than 3 times the maximum value in Fe$_3$Sn$_2$ ($\sigma_{AH}^{int}$~200 Ω$^{-1}$ cm$^{-1}$) [14]. The value of the AHC calculated with density functional theory (DFT) using the Vienna ab-initio simulation package (VASP) [40] and WannierTools [41] package was approximately 507.7 Ω$^{-1}$ cm$^{-1}$, which was close to the experimental value and which further demonstrated that the large AHC was mostly induced by the intrinsic mechanism. Moreover, the large AHC was robustly independent in a wide range of temperatures between 5 and 350 K despite the complex magnetic structure transformation, which indicates that the Berry curvature of the momentum space was very stable. For comparison, the typical physical parameters of Fe$_5$Sn$_3$ and Fe$_3$Sn$_2$ are summarized as shown in Table I. As a result, it was found that the K$_u$ and M$_s$ values of Fe$_5$Sn$_3$ were also larger than those of Fe$_3$Sn$_2$, so they could be a good pair of samples with which to study the correlation between the topological and magnetic properties.

## IV CONCLUSIONS

In summary, a detailed investigation of the magnetic and electrical transport properties of hexagonal ferromagnetic $Fe_5Sn_3$ single crystals is presented in this paper. The MR signs changed at 50 K and the types of majority carriers changed at approximately 275 and 30 K, which was tightly related to the magnetic structure transformation. Moreover, the AHC in $Fe_5Sn_3$ was mostly induced by the intrinsic mechanism and the value of $\sigma_{AH}^{int}$ was approximately 613 $\Omega^{-1}$ cm$^{-1}$, which was more than three times the maximum value in the frustrated kagome magnet $Fe_3Sn_2$ single crystals. More importantly, the large intrinsic AHC turned out to be almost independent of temperatures, indicating that the hexagonal ferromagnetic $Fe_5Sn_3$ single crystal was an excellent candidate for studying the topological features in ferromagnets.

## ACKNOWLAGEMENTS

This work is supported by the National Key R&D Program of China (Grant Nos. 2017YFA0206303), National Natural Science Foundation of China (Nos.11974406 and 11874410).

**Figure caption**

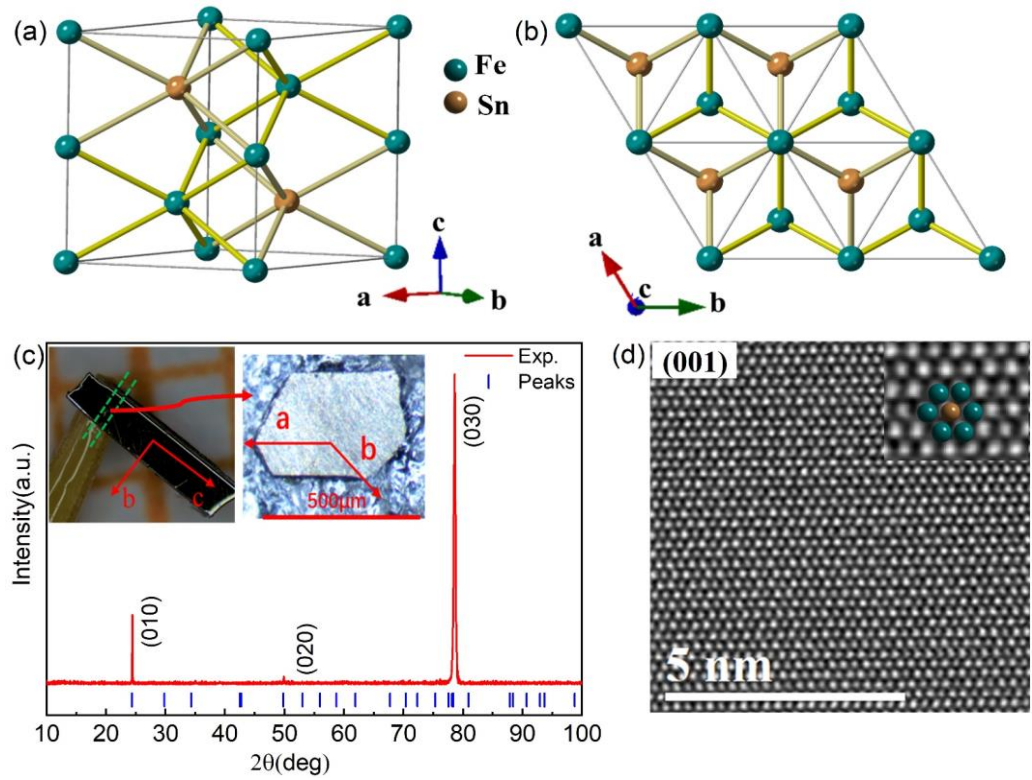

FIG. 1. (a) Side and (b) top views of $Fe_5Sn_3$ crystal structure. (c) X-ray diffraction pattern of bc plane. Inset contains the photographs of a single crystal (left) and its cross-section (right) cut along the green dashed line. (d) High-resolution scanning transmission electron microscopy image of (001) plane.

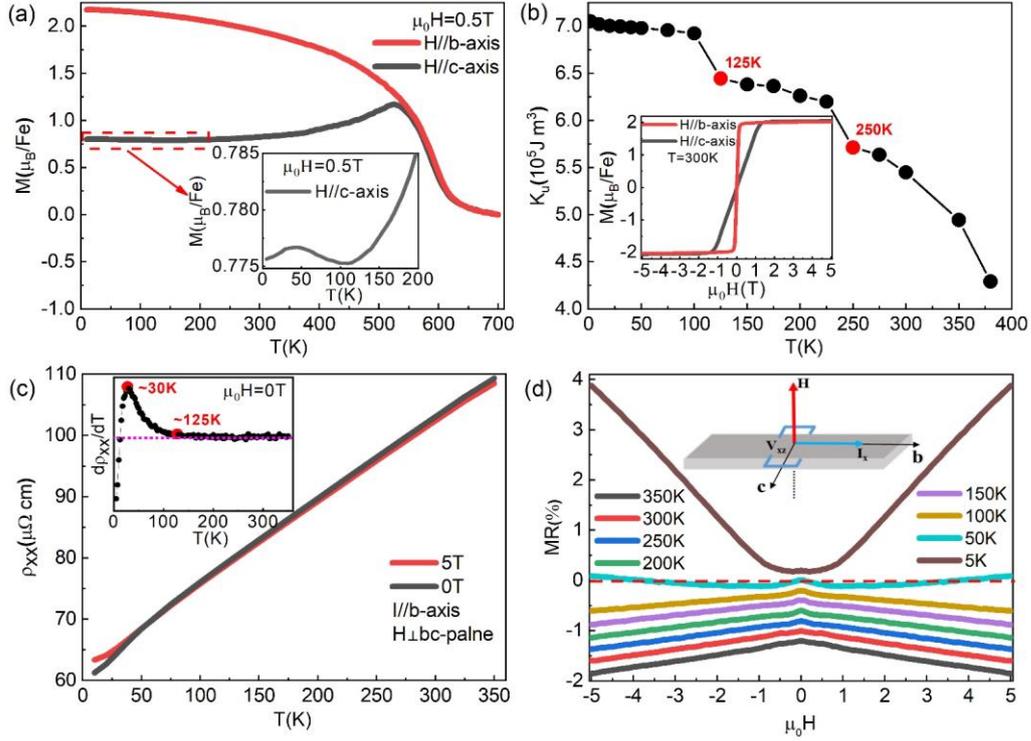

FIG. 2. (a) Zero field cooling (ZFC) curves at 0.5 T along b axis and c axis. Few differences existed between the zero field cooling and field cooling (FC) curves at 0.5 T. To clearly show the M-T curves, only the ZFC curves are shown. Inset shows the low temperature details of the curve with H along the c axis. (b) Temperature dependence of the uniaxial magnetic anisotropy coefficient $K_u$. Inset shows the magnetization curves along the b axis and c axis at 300 K. (c) Resistivity along the b axis for zero field and 5 T. Inset shows the curve of $d\rho_{xx}/dT$ for zero field. (d) Magnetic field dependence of magnetoresistance obtained in the temperature range 5–350 K, with the magnetic fields applied normal to the bc plane and the current along b axis. Red dashed line represents the zero of MR. Inset shows directions of the current and magnetic field in the measurements.

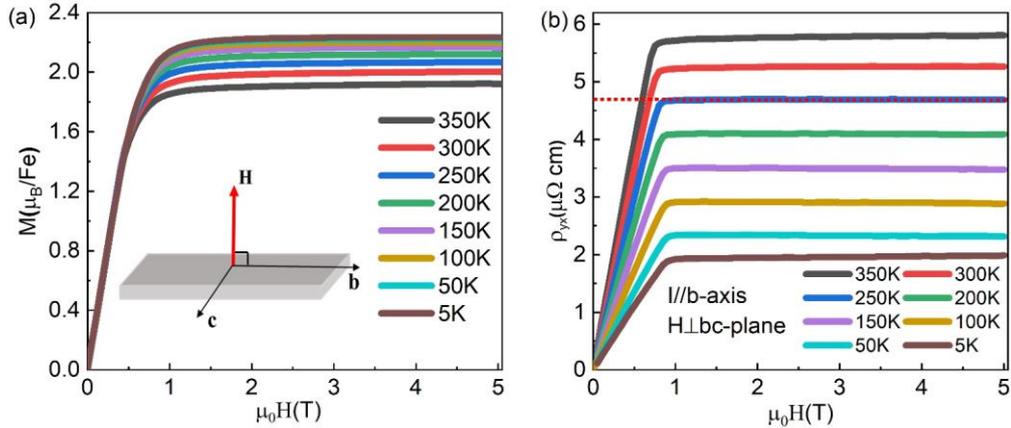

FIG. 3. (a) Magnetization curves at H normal to bc plane. Inset shows the schematic of the magnetic measurement. (b) Hall resistivity $\rho_{yx}$ at current along the b axis and H normal to the bc plane. Red dashed line represents the linear fit of $\rho_{yx}$ at the high field region at 250 K. For clarity, only some data for the temperature are shown. More detailed data are shown in Figs. S3(a) and S3(b) in the Supplementary Material [24].

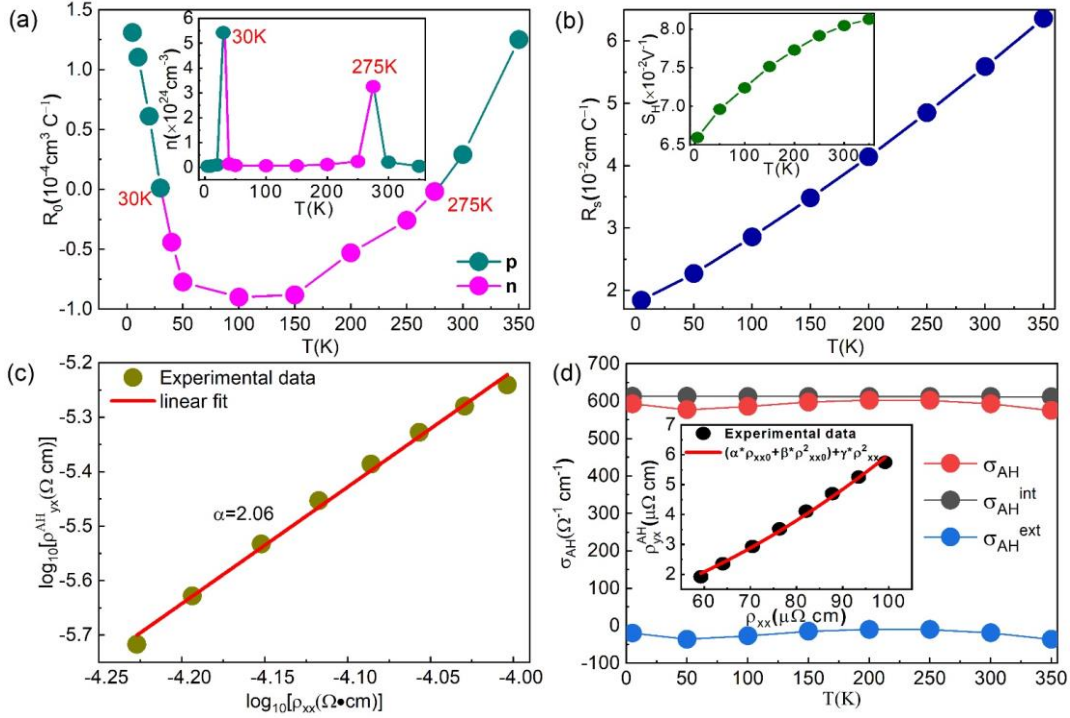

FIG.4. Temperature dependence of (a) ordinary Hall coefficient $R_0$ and (b) anomalous Hall coefficient $R_s$. Inset of (a) shows the current concentration and inset of (b) shows the scaling coefficient $S_H$. (c) Scaling behavior of $\log\rho_{yx}^{AH}$ vs $\log\rho_{xx}$. Red line is the linear fit. (d) Temperature dependence of the anomalous Hall conductance. Inset shows the plot of $\rho_{yx}^{AH}$ vs. $\rho_{xx}$.

## Table

TABLE I. Experimental and calculated values of intrinsic anomalous Hall conductance $\sigma_{AH}^{int-exp}$ and $\sigma_{AH}^{int-cal}$ uniaxial magnetic anisotropy coefficient $K_u$, saturation magnetization $M_s$ and Curie temperature $T_C$ of $Fe_3Sn_2$ [14,39] and $Fe_5Sn_3$.

|  | $\sigma_{AH}^{int-exp}$ ($\Omega^{-1}\cdot cm^{-1}$) | $\sigma_{AH}^{int-cal}$ ($\Omega^{-1}\cdot cm^{-1}$) | $K_u$ ($\times 10^5 J/m^3$) | $M_s$ ($\mu_B$/Fe) | $T_C$ (K) |
|---|---|---|---|---|---|
| $Fe_5Sn_3$ | ~613 | ~507.7 | 5.45 (300 K) | ~2.29 (5 K) | 600 |
| $Fe_3Sn_2$ | ~200 | —— | 3.04 (300 K) | ~1.97 (5 K) | 640 |